\documentclass[%
 aip,
 jap,%
 amsmath,amssymb,
 reprint,%
]{revtex4-2}

\usepackage{graphicx}
\usepackage{dcolumn}
\usepackage{bm}

\usepackage[utf8]{inputenc}
\usepackage[T1]{fontenc}
\usepackage{mathptmx}
\usepackage{xcolor}

\definecolor{cgreen}{HTML}{139F46}

\begin{document}

\preprint{AIP/123-QED}
\title {Perspective: magnetic nanoparticles in theranostic applications}

\author{A. Coene}
\affiliation{Dept. of Electromechanical, Systems and Metal Engineering, Ghent University, 9000 Ghent, Belgium}
\affiliation{Cancer Research Institute Ghent (CRIG), 9000 Ghent, Belgium}
 
\author{J. Leliaert}%
\affiliation{ 
Dept. of Solid State Sciences, Ghent University, 9000 Ghent, Belgium
}%

\date{\today}

\begin{abstract}
Nanomedicine research recently started exploring the combination of therapy and diagnostics, so-called theranostics, as an approach to offer a more flexible, personal and precise care with improved patient outcome. As magnetic nanoparticles show great potential in a variety of diagnostic and therapeutic applications, they are prime candidates to be used in a theranostic platform to realize this vision. This perspective gives an overview of state-of-the-art magnetic imaging techniques and theranostic applications based on magnetic nanoparticles and discusses their opportunities and associated challenges. In order to address these challenges and to exploit these opportunities to the fullest, we discuss three promising research directions.

The first considers the use of novel magnetic field sequences to utilize the rich magnetic dynamics of the particles, allowing a more accurate diagnosis and boosting the performance of many nanoparticle-based applications. Secondly, we introduce the innovative concept of smart theranostics based on feedback mechanisms between the particle applications and their supporting imaging procedure to enhance the performance of both and to allow real-time monitoring of treatment efficiency. Finally, we show the twofold advantage of applying data-driven models to enhance therapy and diagnostics on the one hand, and for handling the platform's large amount of data and associated decision support algorithms, on the other. The latter research track is extended to include hybrid models in which physics-based and data-driven models are combined to overcome challenges of applications with limited data, making the data-driven part understandable, as well as in uncovering unknown nanoparticle dynamics.  

Contrasting other literature, which mainly focuses on developing magnetic nanoparticles with the right characteristics, we put forward advances in magnetic nanoparticle imaging techniques and applications to enable the use of a broader range of magnetic nanoparticles in theranostics. We seek to emphasize the importance of these building blocks as many research opportunities with a very high potential are still left open. Therefore, we encourage researchers to also take these aspects into account to advance theranostic applications of magnetic nanoparticles to real clinical environments. 
\end{abstract}

\maketitle

\section{Introduction}
Research in medicine is to a great extent directed towards developing and improving disease detection and therapy. The use of nanotechnology in medicine, called \textit{nanomedicine}, has been proposed as a solution to overcome intrinsic limitations in conventional approaches and would allow more effective and safer disease diagnosis and treatment \cite{pelaz2017diverse}. Furthermore, nanomedicine could play an important role in personalized treatment approaches, which benefit the treatment of diseases with varying representations among individuals and tackles patient variability in therapeutic responses \cite{tietjen2015nanomedicine}. This is attributed to the unique chemical, physical, and biological properties of the nanomaterials compared to their molecular and bulk counterparts. Of specific relevance in nanomedicine are magnetic nanoparticles (MNPs\footnote{An overview of all abbreviations can be found in the appendix.}) that have the additional benefit that they can be remotely detected and controlled by applying external magnetic fields \cite{wu2019magnetic}. Because of these advantages MNPs have been successfully employed in a broad range of biomedical applications. 

There are many diagnostic applications that benefit from the unique properties of the MNPs. For example, the MNPs' sensitive and specific registration for diagnostic purposes is enabled by the fact that biological tissues are virtually devoid of ferromagnetic materials and that there is zero signal depth attenuation of the magnetic field from biological tissue (contrasting optical and electrical techniques).
These properties are exploited in MNP imaging techniques that directly measure the particles' location and concentration such as magnetic particle imaging (MPI) \cite{yu2017magnetic}, magnetorelaxometry imaging (MRXi) \cite{Liebl2015} and magnetic susceptibility imaging (MSI) \cite{Ficko2015267,calabresi2015alternate}. Because of the particles' small sizes they can easily interact with cells, viruses, genes, and enter virtually every region in the human body. Additionally, the MNPs are highly tune-able\cite{BATLLE2022168594}: by changing their size, composition or synthesis process, desired physical properties and behaviors can be attained such as slow/fast removal from the body, specific reaction to applied magnetic fields, etc. The particle behaviour can be further fine-tuned by coating the particles with specific molecules so they bind to certain entities. This allows ``molecular imaging'' \cite{Debbage2008}, which aims to visualise biological processes by registering specific cells and molecules using nanoparticles. This increases the sensitivity, resolution or image contrast of established imaging techniques in diagnostics. Examples thereof are MRI \cite{laurent2016mri} and adaptations of ultrasound (US), such as magnetomotive US \cite{ersepke2019performance} and magneto-acoustic tomography \cite{Mariappan2016689}. Moreover, radioisotopes or fluorescent molecules can be attached to the MNPs to make them also visible in other imaging modalities, such as SPECT/PET and optical imaging. This is called \textit{multi-modal imaging} and it increases imaging and hence diagnostic accuracy significantly \cite{c4cs00345d}. 

Regarding therapeutic applications based on MNPs, magnetic hyperthermia is considered one of the most promising therapies, especially for cancer treatment \cite{Perigo2015Fundamentals,fratila2019nanomaterials,RubiaRodriguez2021}. In magnetic hyperthermia the MNPs are exposed to an externally applied alternating magnetic field (AMF). This heats up the MNPs and increases the temperature of their surroundings. When the particles are embedded in tumor tissue, this allows for very local heating of the tumours, without harming the healthy cells. Another promising therapy is magnetic drug targeting (MDT) in which the particles are guided towards desired regions or sites in the body by employing magnetic field gradients \cite{al2016magnetic}. When the particles arrive at the desired location, a controlled release of their therapeutic agent (drug, genes, radionuclides, etc.) is possible upon the application of internally or externally applied stimuli such as changes in pH, osmolarity, magnetic field intensity or temperature \cite{schleich2015iron}. This local drug release allows to decrease the total administered dosage, resulting in less systemic side effects.

From previous paragraphs it can be seen that MNPs play an important role in both diagnostic as well as therapeutic applications. This information, together with the fact that the particles can be made multi-functional and have a large flexibility and versatility, show that the particles have great potential to be used in \textit{theranostic} applications \cite{dadfar2019iron,gobbo2015magnetic,zhao2020multifunctional,van2019smart}. Such applications combine diagnostic and therapeutic services and realize a more flexible, precise and noninvasive treatment of patients. The ultimate goal is to develop a platform which combines multiple diagnostic and therapeutic functionalities based on MNPs. Additionally, this platform permits real-time monitoring of disease progress and treatment together with personalized treatment planning. 

In this perspective, we look at recent advances in MNP imaging techniques that are useful in a theranostic setting and we give an overview of state-of-the-art MNP-based theranostic applications. Subsequently, we pinpoint the main opportunities and challenges that lie ahead and provide three research tracks to address these issues while making maximal use of current and future opportunities. Our approach contrasts most other literature concerning theranostic applications. Whereas  these often focus on the requirements of the nanoparticles\cite{zhu2017magnetic,zhao2020multifunctional,pelaz2017diverse,dadfar2019iron}, this perspective draws attention to the applications themselves, the rich underlying magnetisation dynamics, and especially the accompanying imaging technique as equally important research topics to generate progress in this field. 


\section{Recent advances}
\subsection{Towards a theranostic MNP imaging technique}\label{Sec:Towards}
An important cornerstone of theranostic applications is the underlying imaging technique. In this context, a lot of research in the magnetic nanoparticle imaging community has been directed towards improving or extending state-of-the-art MNP imaging techniques as to enable or to enhance the performance of possible theranostic applications. A critical requirement is that the MNP imaging technique should be able to localize multiple types of MNPs (e.g. different MNP types excellent for detection, treatment, etc.) or MNPs in a different state (e.g. MNP with or without therapeutics, MNPs having a specific temperature, etc.). As the technique MPI was already successful at imaging nanoparticles (of the same type and under the same circumstances) at a fast speed while having a fine spatial resolution in the mm range, this technique is a conspicuous candidate to  be made compatible with future theranostic applications. Therefore this section will mainly focus on MPI.

\begin{figure}
\includegraphics[width=\columnwidth]{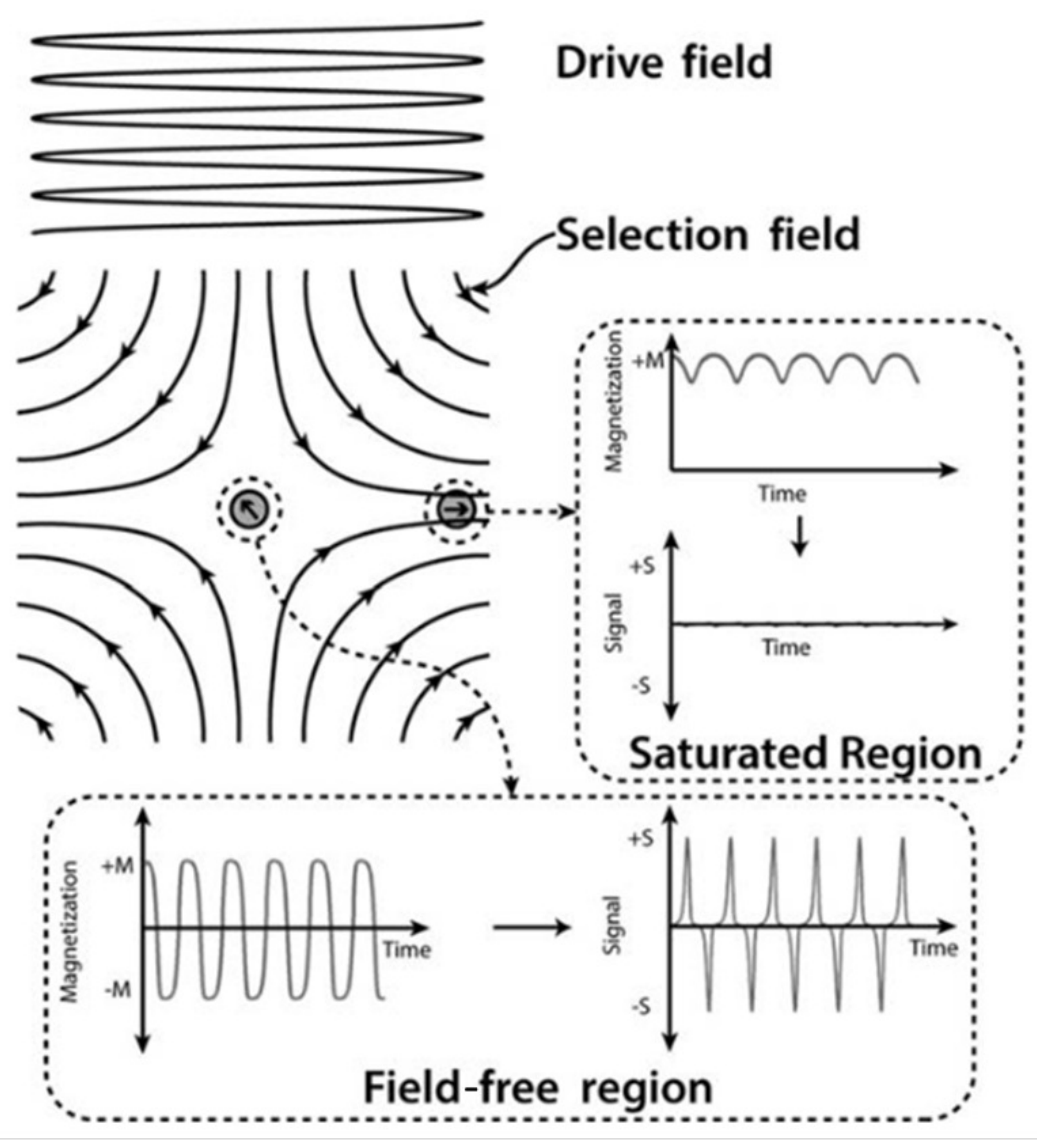}
\caption{\label{fig:MPIprinciples}MPI exploits the non-linear magnetization behavior of the particles by using a selection field that saturates all the particles except those in the FFR. The FFR is shifted over the sample by a drive field, to which the particles in the FFR respond. Reprinted with permission from Dhavalikar \textit{et al.}, Nanomaterials for Magnetic and Optical Hyperthermia Applications, 265 (2019). Copyright 2019 Elsevier.}
\end{figure}

In MPI a static magnetic field, the selection field, containing a field-free region (FFR), is created using magnetic field gradients. In most MPI configurations the FFR is a single point or line, called the field-free point/line. The FFR is shifted over the field of view by a sinusoidally varying magnetic field referred to as the drive field. Only the particles present in the FFR will contribute to the signal as the static magnetic field saturates the MNPs outside the FFR, so they do not respond to the drive field (see Fig. \ref{fig:MPIprinciples}).

In 2015, \citet{rahmer2015first} showed the first experimental separation of up to three particle types \cite{rahmer2015first}. This is often referred to as \textit{multi-color} or \textit{multi-contrast} MPI, as every particle type has its designated color in the final image with the color intensity reflecting the concentration of each particle type. It marks the start of MPI as a promising theranostic imaging technique. This experiment was quickly followed up with the concurrent imaging of MNPs in a vessel together with MNP-coated cardiovascular instrumentation \cite{haegele2016multi} and extended later on with quasi real-time additional steering of the instrumentation using the imaging setup \cite{rahmer2017interactive}. The multi-color approach was also used in a proof-of-concept that showed MPI was able to retrieve the distinct temperatures of two MNP samples, while simultaneously determining their location\cite{stehning2016simultaneous}.

Even though MPI on a single particle type is quantitative (albeit shadowing of lower MNP concentrations by the higher ones can occur, requiring two-step reconstruction algorithms \cite{boberg2021simultaneous}), the ``quantitativeness'' of MPI multi-color imaging is still to be shown. This originates from the fact that MPI signal differences for varying particle types (optimized for MPI performance) are rather small, making their separation more difficult. This could be tackled by increasing the number of MPI measurements by e.g. performing MPI experiments for varying parameters of the applied fields to enhance the information content in the measurements. An example thereof is the work of \citet{Viereck2016} who introduced a dual-frequency acquisition scheme for multi-color MPI to increase the differences in MPI signal for MNP binding state estimation compared to the single frequency approach. By exploiting the frequency-dependent response of the nanoparticles they could qualitatively retrieve the particle's viscosity and amount in 1D. Multi-color imaging was recently shown to be quantitative in discerning between free or internalized MNPs in cells \textit{in vitro} \cite{paysen2020cellular}, due to significant changes in the particles' response when taken up by cells. This result also stresses the fact that changes in particle surroundings or particle - particle interactions can result in image inaccuracies or artefacts \cite{teeman2019intracellular,Lowa2016,them2017magnetic}.

Another challenge of multi-color MPI are the time-consuming calibration measurements that need to be performed for each particle type, environmental condition (viscosity of the surroundings, temperature, etc.) and magnetic field parameter. Such a library of system response functions for different particle types and under varying circumstances is unpractical for theranostic applications. Moreover, this calibration process should be repeated often to correct matters such as system drift or differences in particle response to the applied fields due to changes in particle properties over time. The number of measured calibration functions can be reduced by assuming a linear impact of the varying parameter on the MPI signal, as was recently done in Ref.\cite{salamon2020visualization} to monitor the temperature distribution in liver ablation phantoms. There also exists an alternative image reconstruction approach for MPI  that does not require the measurement of system response functions, referred to as X-space MPI \cite{5728922}. This framework has been used to directly extract a relaxation time image from the MPI signal, in which  the pixels represent the time required to align the particles' magnetic moment with the direction of the applied field \cite{muslu2018calibration}. When the particles can rotate freely, the magnetic moment can change direction by a physical rotation of the particle, referred to as Brownian relaxation. Alternatively, the magnetic moment can alter its orientation within the magnetic core itself, i.e. Néel relaxation. Because these relaxation modes depend on particle properties such as size, surroundings and temperature, the relaxation time image could be used to discriminate between two particle types to obtain a 2D qualitative image of the particle types without performing any calibration measurements. It has also been employed to discern between a single MNP type embedded in 5 viscosity environments in phantoms \cite{utkur2019relaxation}. Although the direct viscosity values could not be retrieved, the change in viscosity could be imaged. Nevertheless, particle-dependent fine-tuning of the magnetic drive field amplitude and frequency for optimal spread between the measured signals of the different viscosity environments was still needed. Recently, a phenomenological model based on the Fokker-Planck equation allowed for the first time in MPI the simultaneous imaging of MNP concentration, viscosity and temperature in a three-line phantom filled with different viscosity MNP samples at different spatial distributions of temperature\cite{PhysRevApplied.16.054005} with the requirement that the MNPs only expressed Brownian relaxation.

In order to be useful in a human setting, the imaging technique should have large field-of-views. At the moment, there are few large field-of-view examples in MPI, as most setups were built for rats and mice. Nevertheless, recently also a human-sized brain scanner for long-term stroke monitoring \cite{graser2019human} and a human torso imager and magnetic targeting setup \cite{rahmer2018remote} were developed, albeit at the cost of a cruder resolution. As MPI directly measures the particles, no anatomical information is available. This information might be necessary to correlate certain findings (e.g. tumor tissue) or to increase imaging accuracy by confirming the particle's location using multiple techniques. Therefore, recently also multi-modal imaging setups were introduced, such as an MPI-MRI \cite{franke2016system}, which has been used for cardiovascular assessments\cite{franke2020hybrid}, and co-registration procedures with CT and/or MRI for anatomic reference images have been successfully performed in e.g. vascular perfusion imaging for example of traumatic brain injury\cite{orendorff2017first} and acute stroke\cite{ludewig2017magnetic}. 

Other MNP imaging techniques, still in their infancy compared to MPI, have in the last years mainly focused on improving their imaging performance\cite{schier2020optimizing,schier2021evaluating,SJOSTRAND20202636,Coene2015c,thrower2019compressed}. Nevertheless, they might also contribute in advancing theranostic applications, as each technique has its distinct advantages. For example, MRXi also developed a multi-color procedure that does allow \textit{quantitative} reconstructions of up to four particle types at larger field-of-view, albeit at a coarser resolution compared to MPI and using an offline reconstruction\cite{coene2017multi}. MRXi has been used to detect ovarian, prostate and breast cancer cell lines and was shown to be several orders of magnitude more sensitive than a mammogram\cite{Hathaway2011}. It also has the advantage that a broader range of particle types can be imaged, still showing significant differences in the measurements. Additionally, the particle imaging technique MSI offers promising up-scaling opportunities towards humans and has already been successful at assessing 2D pharmacokinetics of MNP \textit{in vivo} in rats in real-time \cite{Soares2019Multichannel}. Unfortunately, no MNP quantification \textit{in vivo} has been done with this technique. Combinations of current MNP imaging techniques that could exploit their various advantages and reduce their drawbacks, have not fully been taken advantage of (see Section \ref{Sec:Exploiting}).  

\subsection{MNP-based theranostic applications}
In general, research in MNP imaging techniques has been focused on improving the sensitivity and resolution of the imaging techniques. Because the number of available imaging systems is still small, it follows that clinical applications using these techniques are limited. MPI, however, has recently entered the preclinical research phase with the commercial availability of two MPI systems. Therefore, this section is focused on MPI-enabled theranostic approaches that are now starting to be explored \cite{knopp2017magnetic}. 

The MNPs used in MPI can have long circulation times in the human body allowing for a prolonged registration of the particles, tunable from hours up to several months. This is ideal for vascular imaging, contrasting the nuclear tracers used in PET and SPECT which only last a few hours \cite{zhou2018magnetic}. This advantage has also been exploited in e.g. the quantification of therapeutic effect on tumors\cite{jung2018development} and the tracking of cells in cell-based therapies such as the islet cells in a possible cure of diabetes type 1\cite{wang2018magnetic} and imaging of stem cells\cite{zheng2016quantitative, wang2020artificially} for potential therapy of various diseases such as stroke, myocardial infarction, traumatic brain injury and cancer. Similarly, MRXi has been successfully used in visualising tumours \textit{in vivo} in mice, as well as simultaneously determining the amount of MNPs that arrived at the tumour for determining the efficacy of MNP delivery or in analysing the MNPs circulation time in the body\cite{de2015magnetic}.

A drawback however, is that the injected MNPs, even when targeted with recognition molecules or with the help of magnetic fields, end up in other regions of the body besides the intended ones. This is a key challenge for magnetic hyperthermia experiments as collateral heat damage might occur in healthy tissues. Another challenge faced by magnetic hyperthermia is accurate temperature monitoring as temperature simulations do not grasp the full picture of the complex magnetic dynamics of the MNPs within their \textit{in vivo} environment. In most hyperthermia experiments temperature monitoring is done through fiber-optics probes, but these have the disadvantage that they only measure the temperature at a single point, instead of the complete region under treatment. Moreover, the placement of these probes is highly invasive and challenging for deep-seated tumours \cite{dhavalikar2019image}. A possible alternative, MR thermometry (i.e. measuring the temperature with MRI) is not invasive, but lacks the required accuracy  due to the particles' impact on the MRI signal\cite{tay2018magnetic}. 

MNP imaging techniques could present a solution to these challenges as they allow to locate the MNPs and to monitor the temperature distribution and ablated regions of the tissue, thereby enabling a safe and efficient heating. Moreover, when an AMF, suitable for hyperthermia experiments, is combined with the static field of MPI, heating can be focused in the FFR only, assuring that healthy tissues are spared\cite{chandrasekharan2020using}. Furthermore, the shape of the FFR, and the resulting heating power, can be adapted, thereby enabling tumor-optimized heating \cite{sebastian2019design}. When MNPs with heat-sensitive coatings, carrying drugs, are used, also spatially-controlled drug release might be achieved\cite{fuller2019externally}. Moreover, when combining localized heating with drug-release, a more efficient treatment can be achieved, while simultaneously minimizing systemic toxicity of the drug. 

\begin{figure*}
\includegraphics[width=\textwidth]{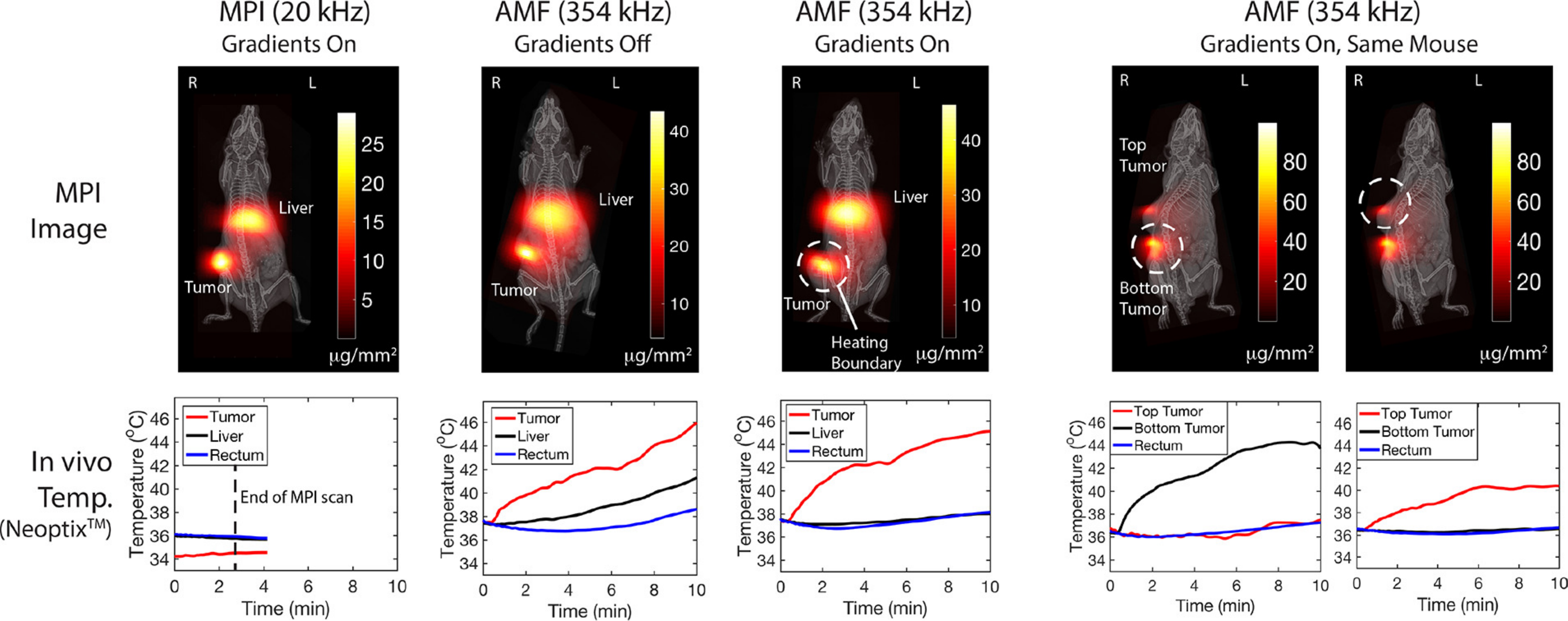}
\caption{\label{fig:localheating} When using MPI to image the MNPs, negligible heating was observed in the mouse thanks to the low frequency drive field (20 kHz) and raster trajectory of the field-free line. When a high-frequency (354 kHz) oscillating field is applied without MPI gradients, all \textit{in vivo} locations with MNPs heat up (thus damaging the healthy liver). When MPI gradients are used, a field-free line is created on the location of the tumor, resulting in heating of the tumor, while the liver is spared. Additionally, a dual tumor mouse demonstrates arbitrary user control of which tumor to heat by centering the field-free line above over the respective tumor. Reprinted with permission from Tay \textit{et al.}, ACS Nano, \textbf{12}, 3699 (2018). Copyright 2018 American Chemical Society.}
\end{figure*}

By fixing the FFR of an MPI setup on the tumor during a magnetic hyperthermia experiment, localized heating of the tumor was realized \textit{in vivo}, while the liver, one centimeter away and also containing MNPs, was unharmed \cite{tay2018magnetic} (see Fig. \ref{fig:localheating}). Additionally, the authors of Ref.\cite{tay2018magnetic} showed the potential of MPI to plan the delivered thermal dose beforehand using the MPI image, which was also supported by other studies\cite{kuboyabu2016magnetic,hensley2017combining}. Monitoring of the temperature using MPI was not done in this study, but could also be included in the future by switching between heating and temperature measurement modes, in which the changes in MPI signal are related to a change in temperature. Other MPI imaging sequences might even allow simultaneous (instead of sequential) heating, imaging and monitoring \cite{wells2020lissajous}. Recently, an add-on module called HYPER, for a commercially available MPI scanner was brought to market that enables controlled heating on millimeter-scales combined with real-time fiber optic temperature monitoring \cite{talebloo2020magnetic}, setting first steps towards full MPI-guided hyperthermia in the future. MPI-guided hyperthermia would greatly reduce the workload of current clinical hyperthermia studies \cite{chandrasekharan2020using}. 

The magnetic field gradients employed in MPI are not only useful for localized heating, but can also be used, together with the generated high magnetic field amplitudes, to apply forces for twisting, bending, pushing and pulling on the MNPs and other magnetic objects. MPI is a flexible magnetic field generator in various directions making it very useful as a general-purpose field applicator for both magnetic targeting and imaging. MRI systems can also switch between imaging and magnetic steering modes and have a number of successful applications under the term \textit{magnetic resonance navigation}\cite{felfoul2016simultaneous,muthana2015directing}, but these systems have smaller gradient fields and less flexibility in field direction compared to MPI\cite{rahmer2018remote}. Additionally, most MNP types result in signal loss in MRI, making their quantification very hard. Therefore MRI is less suitable for MNP-based theranostic applications.

Most MPI-guided targeting applications under development are in the field of interventional medicine for the steering of instrumentation thanks to MPI's excellent performance for quantifying stenosis\cite{Herz2018Magnetic}, balloon catheter tracking\cite{herz2018magneticb} and stent placement monitoring\cite{herz2019magnetic}. Magnetic steering and imaging of catheters when manoeuvring complex vasculature was realized in phantoms\cite{rahmer2017interactive}. In this study magnetic forces could be applied on the catheter to direct it into the desired vessel branch by adding a soft magnetic sphere on the catheter's tip. Catheter visualisation was possible by attaching a small compartment containing a MNP powder and the vasculature was imaged simultaneously by filling the phantom with a diluted MNP suspension. Not only catheters can be manipulated, \citet{nothnagel2016steering} showed how a regular MPI imager can generate forces of variable strength and direction for small softmagnetic spheres and needles and allows automatic control of these devices towards a desired position in phantoms based on MPI image information \cite{nothnagel2016steering}. Also a millimeter-sized magnetically coated swimmer was successfully steered\cite{BAKENECKER2019495}. Recently, a clinical scale MPI actuator was built that performs magnetic targeting on human-sized objects and was used to steer magnetic drills using high torques through gel and tissue samples on specific trajectories\cite{rahmer2018remote}. This latter approach might be useful in minimally invasive surgery and for the manipulation of microrobots, as was recently successfully demonstrated by the navigation of a magnetic micro-robot through a human cerebral aneurysm phantom \cite{bakenecker2021navigation}. 

Next to the steering of magnetic devices, also MNPs can be manipulated using MPI. As these particles are generally significantly smaller, the generated forces also decrease severely, making nanoparticle manipulation not straightforward. Especially, when the MNPs need to be steered against blood streams or over biological barriers. Recently, however \citet{Griese2020Simultaneous} showed the possibility of navigating MNPs towards a narrowed or blocked vessel by stenosis in a bifurcation phantom, while also visualising the particles' accumulation on the stenosis \cite{Griese2020Simultaneous}. A mixture of micro- and nanoparticles was used to have both adequate imaging as well as targeting performance. Several phantom inflow speeds and stenosis degrees were analysed and the results indicate that it might be feasible to resolve a complete stenosis in medium-sized arteries using drug-loaded MNPs. Intravenous injection of the drug without MNPs would never reach the stenosis, because the drug is drawn towards the unblocked branch, hence traditionally, a catheter is employed to clear the clot. MNP-targeting might remove the need of invasive procedures and might reduce the required drug dosage. The experiments without stenosis also showed that MPI probably could be used to move particles in larger branches, were blood flow is increased, such as from the aorta to the carotid artery. Moreover, in small arteries it was possible to fix particles against the wall, which is of importance for drug targeting applications.


\section{Challenges \& opportunities} 
The previous sections show that MNPs are promising for a plethora of theranostic applications. Nevertheless, there still are some challenges that need to be overcome in order to reach the MNPs' full theranostic potential. In this section, several opportunities that might improve the MNPs performance in theranostic applications are discussed together with the corresponding challenges.

\subsection{Exploiting complex magnetic particle dynamics}\label{Sec:Exploiting}
Focus over the last years has been foremost on integrating MPI with applications such as magnetic hyperthermia or drug targeting or improving application/imaging performance. The latter was achieved mainly by particle and hardware enhancements and innovative reconstruction algorithms. One aspect that has remained underexposed is the information content that can be retrieved from novel magnetic field sequences. In general, the magnetic fields employed in the applications or imaging technique did not undergo many adaptations, while significant improvements can be realized from further exploitation of the rich MNP magnetisation dynamics. 

For example, the main MNP imaging techniques (e.g. MPI, MRXi, MSI) each use a different range of magnetization dynamics of the MNPs. For instance, in MRXi the delayed response of the MNPs to a suddenly removed magnetic field pulse is measured, which can be directly related to their relaxation time constant (see Section \ref{Sec:Towards}). This information might be of value for example in MPI experiments, to assess the lag in the particles' response to the FFR, as this causes blurring effects that reduce the achievable resolution (see Section \ref{Sec:Towards}). \citet{tay2019pulsed} introduced \textit{pulsed} MPI, a combination of MRXi and MPI as a first step in this direction, and showed how it significantly improves the resolution of MPI in 2D line phantom experiments\cite{tay2019pulsed}. Subsequently, new hardware concepts were introduced to further facilitate \textit{pulsed} MPI \cite{viereck2020initial}. Additionally, the multi-color procedures for MRXi, developed in Ref. \cite{coene2017multi}, could be applied to \textit{pulsed} MPI, to further improve multi-color imaging. Moreover, the techniques that allow to estimate the relaxation time constant in MPI (see Sec. \ref{Sec:Towards}) can be used as additional information to increase the resolution of multi-color imaging further \cite{muslu2018calibration,PhysRevApplied.16.054005,rahmer2015first}. 
Additionally, the differences between the obtained relaxation time constants from different approaches can be used to quantitatively assess the impact of the used magnetic fields in every approach. Indeed, changes in relaxation time might be of value in measuring the degree of dipolar interactions and other field dependencies in the sample, which can impact application and imaging performance\cite{Leliaert2014,them2017magnetic,Leliaert2017Interpreting,D1NA00463H}. In a recently developed MNP characterization technique called \textit{thermal noise magnetometry}\cite{leliaert2015thermal}, this idea is carried to the extreme, as it uniquely measures the magnetic noise originating from the thermally induced fluctuations in the particles' magnetization in the absence of any externally applied field. It is shown that the obtained noise spectra carry similar information as other characterization techniques\cite{leliaert2017complementarity}. Additionally, this technique offers a unique and sensitive window into the aggregation state of the particles thanks to the underlying scaling laws of some of the noise properties with respect to particle clustering. 
This can be exploited in applications, e.g., to monitor the particle aggregation in immunoassays with increased accuracy\cite{everaert2021noise}.

Turning our attention again to nonzero fields, \citet{Jaufenthaler2020OPM} showed how the relaxation time was already affected by small DC fields ($<$ 100 $\mu$T) and their applied direction  \cite{Jaufenthaler2020OPM}. Similarly, the measured phase lag in MSI is also related to these relaxation time constants and could provide information on changes in particle dynamics and interactions for the case of small AMFs.  Taking these effects into account would also result in an improved characterisation of the MNPs as this is often based on measuring the relaxation time constants without consideration for the impact of magnetic field properties and MNP interactions\cite{bogren2015classification,liebl2021magnetic}. 

One could envision a general imaging framework based on combinations of MPI, MRXi and MSI. Given the type of particles and experimental conditions, one of them or a combination thereof can be used to reach the desired specifications with respect to imaging resolution, measurement time and field-of-view. Starting from sequences that are initially combinations of state-of-the-art imaging sequences, a gradual extension toward sequences containing both temporally and spatially varying magnetic fields, such as pulsed, gradient, static, and AC fields can be made. Such a system (i.e. a general magnetic field applicator) would also facilitate the merging of different applications to create a theranostic platform. For example, the required DC gradients for magnetic targeting and the AC fields for magnetic hyperthermia can be combined. Moreover, a more thorough characterisation of the MNPs can be performed by measuring their response to this versatile set of magnetic fields, 
allowing to exploit the strengths and to mitigate the weaknesses of each individual technique.
We recently demonstrated this advantage by developing a fitting procedure based on a combination of different magnetic characterisation measurements, each sensitive to a certain range of magnetic dynamics\cite{karpavivcius2021advanced}.
This way we could accurately determine the properties of magnetic nanoflowers which have the remarkable feature that they display dynamics over a timescale spanning 5 orders of magnitude, making them very useful for a broad range of applications. The obtained results of the combined fit were not only consistent with each other, i.e. each of the individual characterisation data sets, but provided more detailed information (e.g. more finely resolved peaks in the MNP size distribution) in the range accessible to multiple methods. 
An accurate characterisation is crucial to fine-tune platform settings towards specific MNP sizes, agglomerates, particle interactions or relaxation rates, boosting the applications' performance. 

\begin{figure*}
\includegraphics[width=\textwidth]{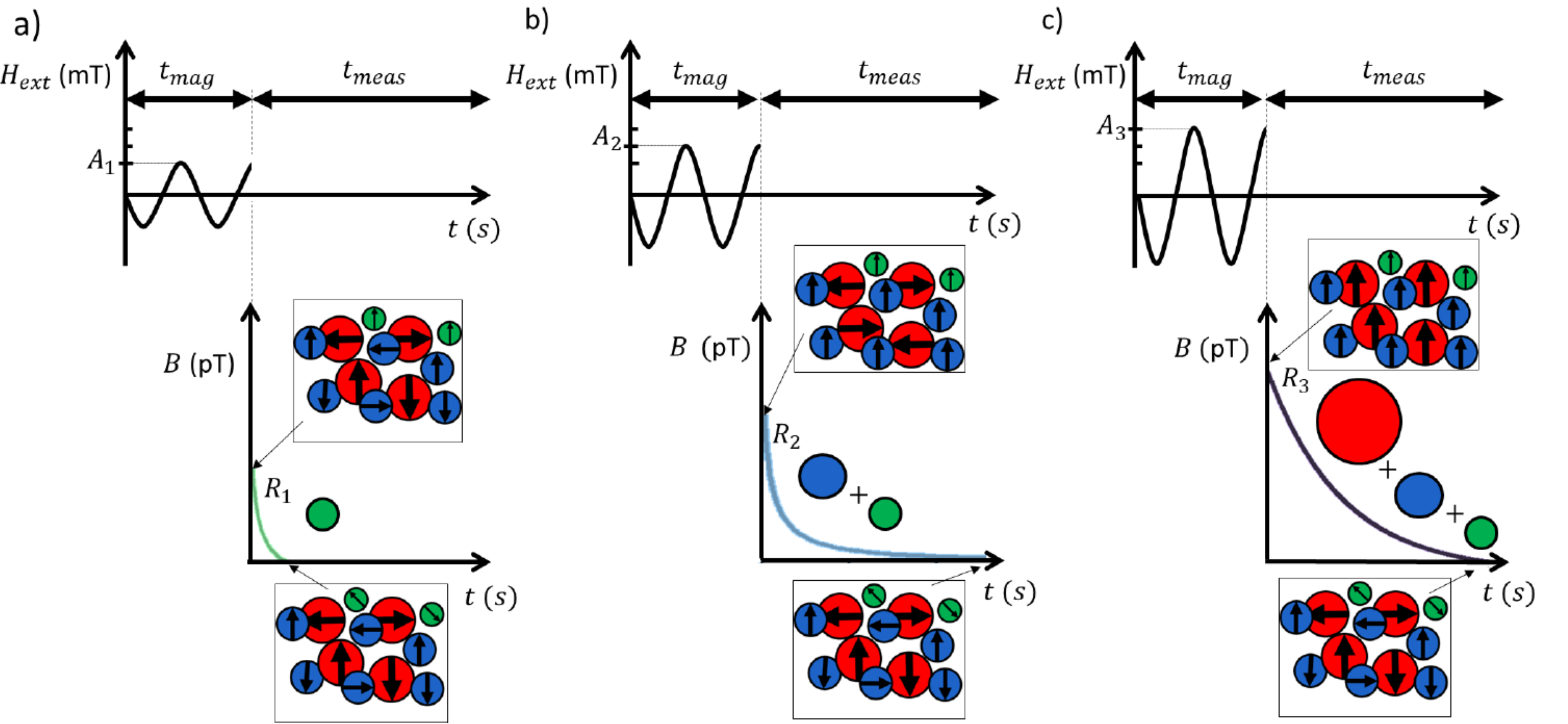}
\caption{\label{fig:selectiveactivation} Using a pulsed sine instead of a conventional constant pulse in MRX, it becomes possible to activate specific subsets of MNPs. By changing the amplitude of the pulse, a different number of particle types respond to the pulse. a) When a sine pulse with amplitude $A_1$ is applied, only the green particle type fulfills the requirements and hence aligns to the pulse. When the pulse is removed, the sensors only measure the decaying magnetic signal of this particle type as the others are still randomly oriented. b) When a sine pulse with amplitude $A_2$ is applied both the blue and green particle types respond and hence both particle types contribute to the measurements. c) When a sine pulse with amplitude $A_3$ is employed all the particle types are activated and the decaying signal contains the signals of all the particles. By subtracting subsequent measurements, individual MNP information can be retrieved. Reprinted from Coene \textit{et al.}, Sensors, \textbf{20}, 3882 (2020).}
\end{figure*}

In this context, we recently improved the characterization performance of MRX by using a pulsed sinusoidally or circularly polarized AC field instead of a DC pulse\cite{coene2020simultaneous}. Intriguingly, the response of the particles to such fields is very different depending on whether the field amplitude is larger or smaller than a, particle-dependent, critical value\cite{usadel2015}. By increasing the field's amplitude, more particles fit this requirement, and show a dynamical response that can be picked up by the sensors (illustrated by Fig. \ref{fig:selectiveactivation}). By subtracting subsequent MRX measurements, only the response of a specific particle type can be obtained. This way, it became possible for the first time to accurately determine the size distribution of the particles, while simultaneously retrieving their magnetic coercivity. These properties are for example of interest for magnetic hyperthermia, as the size and coercivity of the particles determine the optimal applied field frequency and the resulting heat generation. Inclusion of such a pulse in magnetic hyperthermia experiments would allow for real-time characterization and fine-tuning of the magnetic field towards an optimal  heat generation. Such an approach is also interesting to realize multi-color imaging without the need to deconvolute the particles' responses. This contrasts previous approaches which always required a model or software deconvolution approach to discern the specific particle type information in postprocessing\cite{coene2017multi,rahmer2015first,stehning2016simultaneous,muslu2018calibration}. Instead, the presented approach allows the different particle types to be independently activated, thereby enabling  the \textit{direct} imaging of a specific particle type without any pre- or postprocessing or calibration measurements of the different particle types. This approach can readily be applied to MRX imaging or \textit{pulsed} MPI, by replacing the currently used DC pulses by sinusoidally varying field pulses.
The sequence could also be included in existing imaging sequences to increase the knowledge of particle properties (and hence possibly optimize experiment settings) and/or to selectively activate a set of particle types. 
Note that it is not necessarily needed to subtract subsequent measurements, as the differences between the set of activated particle types between measurements can be interpreted using a model-based reconstruction. 

The application of various magnetic field sequences does not only facilitate the merger of various applications, it also allows to improve their effectiveness. As shown in the previous paragraph, magnetic hyperthermia relies on an accurate characterisation of the MNPs to fine-tune the magnetic field frequency and amplitude. To this end, it is essential to be able to assess the particle's state given the observed discrepancy between \textit{in vivo} and \textit{in vitro} heating performance, attributed to particle clustering inside the cell\cite{cabrera2018dynamical} or to a change in the particle's mobility due to e.g. proteins attaching to the MNPs \cite{raouf2020review}. How the magnetic dynamics are altered \textit{in vivo} can be exposed by these magnetic field sequences and would allow for more targeted heating optimisation by adapting the therapy parameters \textit{in operando}. Also useful in this context is the fact that heat generation can be improved by using more complex excitation schemes instead of the commonly employed linearly polarized sinusoidal field. \citet{dejardin2017effect} suggest that the addition of a small DC magnetic field impacts the heat generation depending on the particle interactions \cite{dejardin2017effect}. Indeed, the DC magnetic field might induce specific MNP configurations such as chains or clusters which impact the MNPs' performance in several applications\cite{Leliaert2014,Laslett2015,them2017magnetic,abu2020theory,wu2017magnetic,lowa2016concentration}. The influence of dipolar interactions and DC fields was also recently experimentally observed in AC susceptibility and hysteresis experiments, indicating an improved heating for magnetic hyperthermia\cite{onodera2020dynamic,shi2019enhanced,chen2020tuning}.  A small body of work used rotating magnetic fields in hyperthermia experiments\cite{bekovic2018magnetic}. In numerical simulations it was shown that  enhanced heating performance (by 30 to 40 \% compared to standard AMF hyperthermia) can be realized for both strongly or weakly interacting MNPs, partially attributed to the fact that a wider diameter range of the MNPs can contribute to the heating\cite{usov2020heating}.  Previous studies indicate that a significant improvement can be achieved with  small alterations to the applied magnetic field, holding promise for more complex spatial and temporal varying field sequences. As another example of the versatility in responses that can be achieved with complex field excitations, \citet{wells2020lissajous} demonstrated that the Lissajous imaging sequence in MPI can also be used for hyperthermia purposes which would allow possible simultaneous heating and imaging \cite{wells2020lissajous}. The analysis of the multitude of fields currently under study is therefore also critical from a safety viewpoint to assure no undesired heating occurs in other applications.  

Also other theranostic applications, such as MDT, could benefit from complex magnetic field patterns.  MDT experiments and simulations so far have only focused on  magnetic field gradients (with the exception of MPI-based studies). Nevertheless, combinations of fields could induce/break up MNP clusters or chains, allowing for larger/smaller forces acting on the MNP. Also the timing of these fields is important, as it was observed for example that chain length and corresponding targeting speed of certain particle types is related to field application time\cite{BENHAL2019187}. The fields can also induce particle shape variations, impacting drug release. Similarly, magnetic field sequences can be used to exploit multi-color information to e.g.visualise the release rate/state of a drug.

One step further down this road is the modelling of the magnetisation responses to these complex magnetic field patterns. This can be done by starting from the underlying micromagnetic dynamics described by the stochastic Landau-Lifshitz-Gilbert equation. This is a computationally intensive endeavor\cite{tomorrow}. Especially for structures displaying complex internal magnetization dynamics, such as the dynamics of magnetic vortices in a nanodisk as an alternative to magnetic nanoparticles in hyperthermia\cite{manzin2021micromagnetic}, GPU-accelerated micromagnetic software\cite{manzinGPU,leliaert2018fast} is an almost indispensable tool. 

\begin{figure}
\includegraphics[width=\columnwidth]{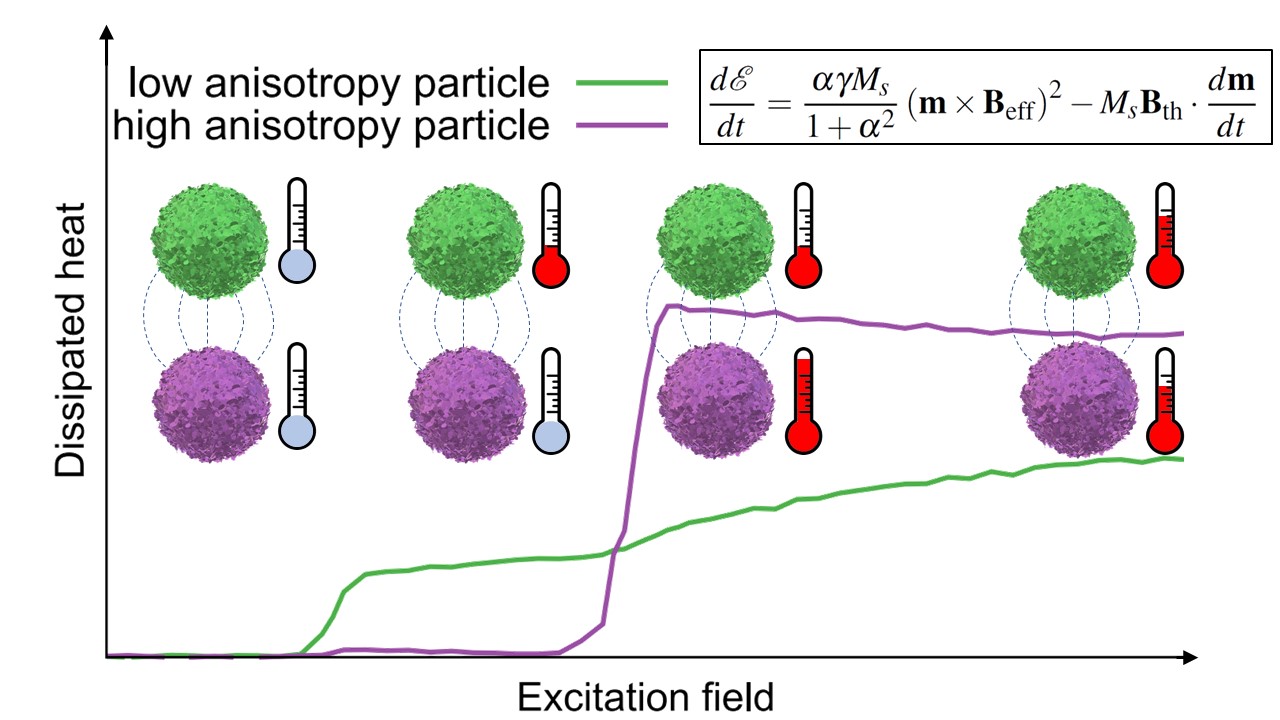}
\caption{\label{fig:homogeneous} 
The equation shown at the top right can be used to resolve the individual heat dissipation of interacting
nanoparticles at nonzero temperature. The green and purple lines show the individual heat dissipated in each of two interacting magnetic nanoparticles with different anisotropy constant. The heating tends to become more uniformly distributed for larger fields. Reprinted with permission from Leliaert \textit{et al}., Nanoscale, \textbf{13}, 14734–14744 (2021).} 
\end{figure}

For typical magnetic nanoparticles the numerical problem can be simplified and calculated with traditional CPU-codes by assuming that the particles are uniformly magnetized\cite{leliaert2015vinamax,laslett2018magpy}. Within such a macrospin approach, all relevant physics, like the emergent switching times under the complex conditions of interacting nanoparticles at nonzero temperatures, are automatically included. Otherwise they have to be accounted for using statistical models\cite{hovorka_chain}, making the macrospin approach very versatile. Furthermore, such numerical approaches allow to access quantities that are very difficult to assess experimentally. One example is the heat generated by each individual particle in an ensemble\cite{oksana, lel-nanoscale}. To illustrate the relation of such numerical results to real-world applications, a recent study by one of the authors\cite{lel-nanoscale} showed that  a more homogeneous heat distribution could be obtained at larger applied field intensities despite a constant total amount of generated heat (see Fig. \ref{fig:homogeneous}). This offers a pathway to mitigating the long-standing problem of hot-spots in hyperthermia. 

CPU-codes allow to deepen our insight in the magnetization dynamics of modestly-sized ensembles of nanoparticles. However, when performing multi-physics simulations, e.g., in which the mechanical rotations and magnetization dynamics are coupled, of large ensembles of particles, the use of GPU-accelerated software again is warranted. For instance, \citet{cabrera2017unraveling},
consider the influence of the viscosity of the medium on the low frequency hysteresis loops of magnetic colloids
consisting of iron oxide or cobalt ferrite nanocubes of different sizes and chemical composition. Their results show that such numerical work can help to understand the sensitivity of the heat dissipation towards viscosity, thereby
allowing to engineer particles that are as affected as little as possible by changes in their environment. This constitutes a step towards maximising the \textit{in vivo} heat dissipation from magnetic nanoparticles, where they are subjected to the complex dynamic environment of living tissue.

To realize the vision of a theranostic platform, a significant challenge needs to be addressed related to the hardware aspects of an all-round magnetic field generator with respect to appropriate power and system design and in some cases shielding of magnetometers. As a final point we would like to stress that care needs to be taken to not excessively lengthen measurement time by introducing a plethora of consecutive magnetic field patterns. Consequently,  research should also be conducted towards optimized magnetic field sequences that both maximize the desired information content in the measurement signals (and/or application performance), and minimize the measurement (and/or treatment) time. 

\subsection{Smart theranostics}\label{Sec:Smart}
Up until now, focus in literature has mostly been on the hardware side of integrating the imaging modality with its supporting application(s). Current MNP-based theranostic applications at most make use of an imaging model that relates the measurement signal to a spatial particle distribution, even though many models of MNP-based applications with various complexities are available \cite{raouf2020review,al2016magnetic}. Adding these application models next to the imaging model would result in a significant safety and efficiency gain as e.g. the generated temperature and the displacement of the particles could be estimated. Building on the concepts introduced in  Section \ref{Sec:Exploiting}, which aimed to increase the information content in the measurement signals, MNP-based theranostics can be made smart by introducing these models in the platforms. There they exploit the information content in the measurement signals and provide the system with the ability to anticipate and to intelligently use this information. 

Even more groundbreaking results are to be expected by the combined use of these models in which the generated knowledge by the individual models can be passed on to the other models in the system as \textit{a priori} information and also be used to provide feedback to one another. In this case, for example, predictions on locations of the particles from the imaging model can be fed to an application model of magnetic hyperthermia or targeting to improve its performance. Vice versa, a predicted displacement of the particles by a magnetic targeting model can be used in the imaging model, or an estimated temperature increase can be taken into account by the imaging model to foresee a change in particle response, while remaining at the same location. 

\begin{figure*}
\includegraphics[width=\textwidth]{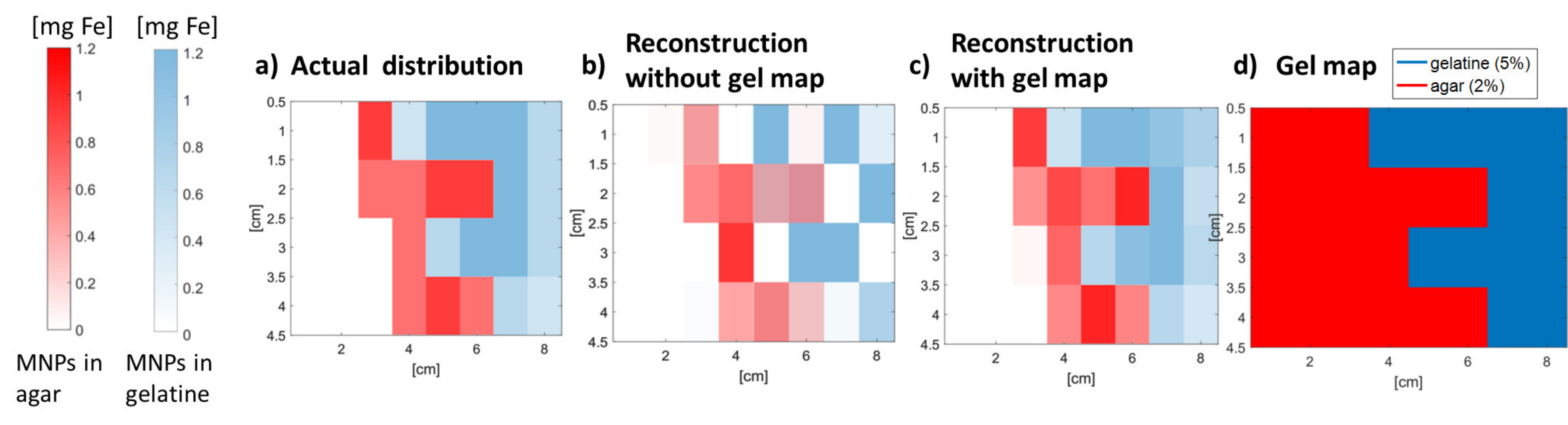}
\caption{\label{fig:mrxi_tissue} When MNPs are embedded in 2\% agar and 5\% gelatine respectively, the resulting MRXi signals vary only slightly. a) In this phantom MNPs embedded in agar (red) and in gelatine (blue) are also in close spatial proximity, which further complicates the separation of their signals. b) The reconstruction of the phantom from panel a) using MRXi, shows that it is hard to differentiate between the particles in both environments. c) MRXi reconstruction performance can be significantly improved by exploiting the information on tissue properties from other modalities, such as a viscosity map as depicted in panel d). By limiting the reconstruction to each respective environment, an accurate representation of the particles can be retrieved. d) The gel map presents the spatial variation of both gel types (in red for 2\% agar, blue for 5\% gelatine), but has no information on particle locations, and is used in defining the spatial reconstruction boundaries for each signal type.} \end{figure*}

Such an approach could reduce the issue of uncontrollable temperature rises in magnetic hyperthermia\cite{etemadimagnetic}, where for example in clinical trials temperatures of 82 $ ^{\circ}$C were measured in the treatment of patients having glioblastoma (a very aggressive brain cancer)\cite{maier2011efficacy}. In this case, a magnetic hyperthermia model could be employed which anticipates the temperature increase starting from the predicted MNP locations by the imaging model. In turn this response could be coupled to a magnetic field model, which adapts the currents in the coils, to ensure suitable magnetic field parameters to allow safe hyperthermia. As an additional check, the techniques from multi-color MPI could be exploited to estimate the temperature of the sample. Similarly, in recent work, we showed the advantages of linking predictions of a magnetic nanoparticle targeting model to a control algorithm of the coils to generate specific magnetic forces on the particles \cite{Durme2021Model}. This approach made it possible to mitigate dispersion of the nanoparticles and to bring them to their desired location, following the shortest path under minimal coil energy requirements and keeping in mind the limitations of the coils. Figure \ref{fig:mrxi_tissue} illustrates another example of the positive impact of feedback mechanisms. In this case information on spatial tissue properties (from e.g. an MRI or multi-color MPI/MRXi techniques) is used to improve MNP image reconstruction results. Simulations for particles embedded in 2\% agar (red) and 5\% gelatine (blue) gel environments respectively show a similar MRX signal. When both environments are close to each other and filled with MNPs (Fig. \ref{fig:mrxi_tissue}a)), it becomes very difficult to separate both particle signals, deteriorating the reconstruction quality (\ref{fig:mrxi_tissue}b)). MRXi reconstruction results can however be significantly improved (\ref{fig:mrxi_tissue}c)) by limiting the reconstruction region of a specific particle response (i.e. agar or gelatine) based on information of the gels' location obtained from the map (Fig. \ref{fig:mrxi_tissue}d)) .

The innovative use of feedback mechanisms between the models is very promising. Nevertheless, it needs to be carefully analysed which feedback aspects should be included as not to worsen performance or to make the system unstable when inaccurate information is passed on. One way to resolve this issue is to analyze multiple information channels on the same parameter. For example MPI multi-color techniques could be exploited to estimate the temperature which can be compared to temperature estimations of the magnetic hyperthermia model. When both are in agreement with each other the feedback can be used in other models, and if this is not the case, other information channels of this parameter could be exploited to improve the accuracy of the feedback or, when no additional channels are available, the feedback can be ignored. 

Another point of attention concerns the ability of having real-time imaging and application information to ensure the system's continuous safety and efficiency. There is a trade-off between the accuracy and computation speed of results obtained, e.g. slow but accurate complex models and fast, but less reliable ones.
It should therefore be investigated how models of varying complexity can interact with each other to ensure sufficiently accurate results without jeopardizing the feasibility of real-time decision making. 
Such modeling approaches with adaptive complexity are not only useful for theranostic applications but could possibly also push forward stand-alone MNP-based applications towards real-time assessments.

\begin{figure}
\includegraphics[width=\columnwidth]{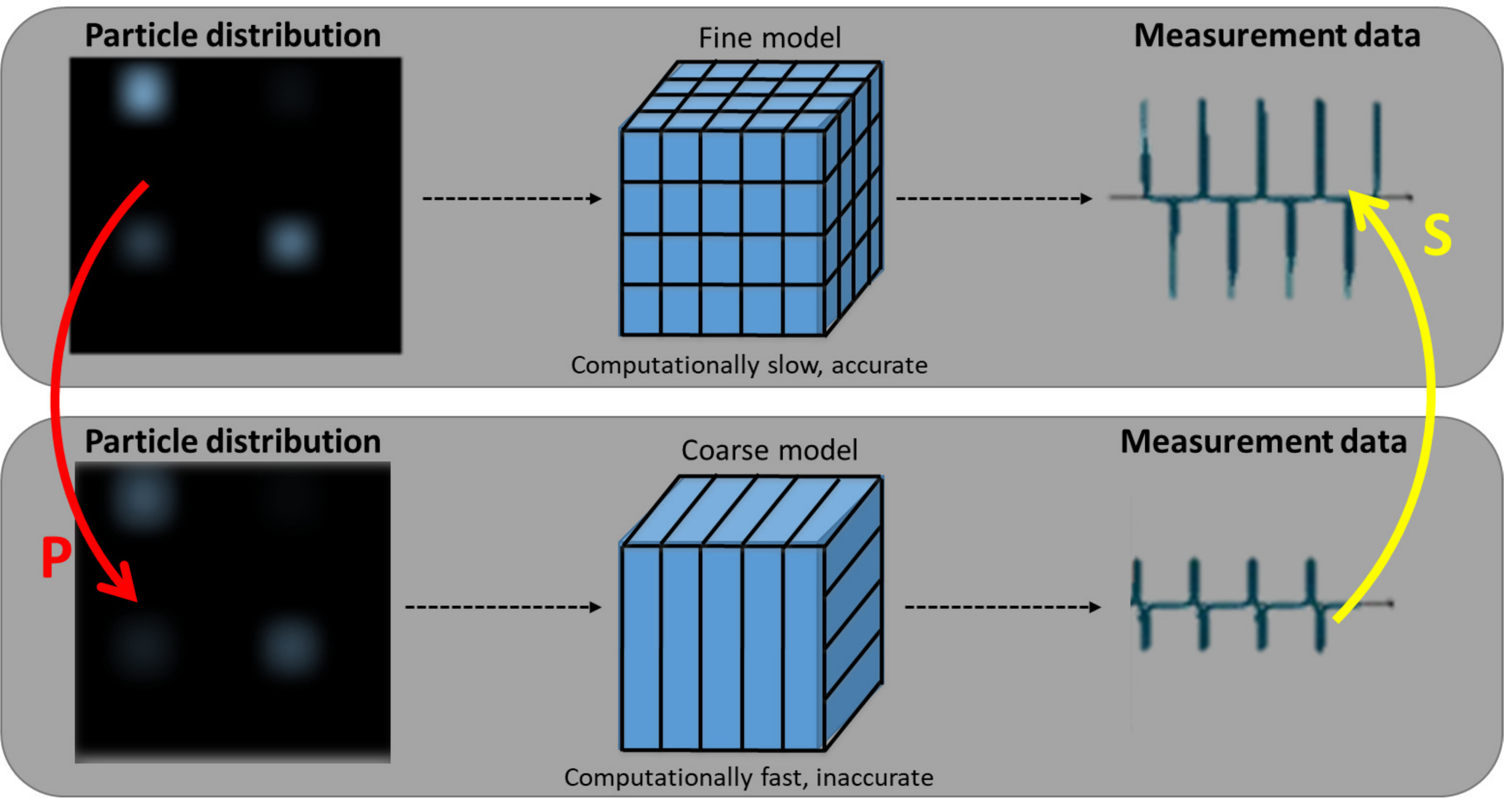}
\caption{\label{fig:varyingcomplexity} An example of how RPM methods can be used in the case of imaging to determine how models with varying complexity need to interact with each other to ensure accurate and real-time imaging. In RPM methods mappings between the inputs ($P$) and outputs ($S$) of a fine and coarse model are optimized so that a surrogate model is found with an accuracy similar to the fine model and a speed similar to the coarse model.} 
\end{figure}

In MNP imaging, a few hybrid modelling examples can be found with the aim to reduce the measurement time of the system response function by combining an analytical model with measurements\cite{19335923} or by merging different measurement sets \cite{von2020efficient} and also model transformations have been suggested for reducing model complexity\cite{7067486,Coene2015c,baumgarten2015plane}. Unfortunately, up until now interacting models with varying complexity have not been pursued. One possible way to include this interaction is through two-level (or more) response and parameter mapping-based (RPM) methods\cite{crevecoeur2012convergence}. Here, use is made of both a fine, accurate model and a coarser model with less accuracy. Through finding the solution (e.g. the particle distribution in imaging), the coarser model is iteratively refined until it has a similar fidelity as the fine model, but with the advantage of finding solutions at the speed similar to the coarser model. In retrieving this surrogate model, only a limited number of evaluations with the fine model are required to improve the surrogate model, while the surrogate model is used to explore the possible solutions. The surrogate model is enhanced by optimizing the mappings between the input ($P$) and output ($S$) space of the coarse and fine model, see Fig. \ref{fig:varyingcomplexity}. This figure shows an example of how RPM methods could be used to provide an answer on how a fine and coarse model should interact to enable real-time and accurate imaging. The coarser model could for example be a low-resolution system function or a simple analytical formulation, while the fine model could be a high-resolution system function.

\subsection{Introducing (hybrid) data-driven models in theranostics}
Exploiting the particle dynamics (Section \ref{Sec:Exploiting}) and the use of model-based feedback in theranostics (Section \ref{Sec:Smart}) inevitably go hand in hand with a significant increase in the types and amount of data . How to cope with the varying data aspects and to utilize the available information to the fullest is touched upon in this section.
            
Most models used in magnetic nanoparticle applications are based on physics\cite{etemadimagnetic,al2016magnetic,BENHAL2019187,5728922}. Generally, these models are able to describe the overall dynamics of the system well for a broad range of parameters and are robust against uncertainties in the inputs. However, sometimes the predicted output deviates from the actual result due to invalid assumptions or simplifications of the model. Alternatively, so-called \textit{black-box models} can be built which are completely driven by data without any knowledge of the underlying physics. One example of this approach is the use of ``system response functions'' in MPI and MRXi \cite{rahmer2015first,coene2017multi} (see Section \ref{Sec:Towards}). Deep learning (DL) techniques\cite{Goodfellow-et-al-2016} are commonly used in other research fields as black-box models. The term `deep' refers to the large number of transformations performed in a layered fashion on the data to transform the input to the desired output. These transformations are for example projections, translations or nonlinear operations. The respective parameters of each transformation are learned by providing the DL model with a training set consisting of true input (e.g. MPI measurement signal) and output (e.g. corresponding spatial MNP distribution) pairs. Contrasting physics-based models, DL models match measurements very well when the measurements are under similar experimental conditions as the measurements that were used in training (e.g. similar temperatures, particles, etc.), however once outside the training region,  DL models typically perform less well. This contrasts physics-based models which are more robust towards measurements outside the training region as compared to DL models.

DL models that are commonly employed and could be of specific interest for MNP-based theranostics are recurrent neural networks, since they capture the nonlinear dynamics in time series\cite{ogunmolu2016nonlinear} and
are, therefore, perfectly suited to learn the MNP dynamics occurring in applications. Nevertheless, recurrent neural networks have not yet been applied in our field. Also of value are convolutional neural networks\cite{mccann2017convolutional}, which have been successfully used in data classification, pattern recognition and inverse problems such as image reconstruction. Only a very recent attempt of applying convolutional networks in MNP-based applications by \citet{baltruschat20203d}
can be found. In this work, super-resolution convolutional neural networks are employed to retrieve a high-resolution system function to be used in MPI reconstruction starting from measurements of a low-resolution system function. This approach reduces system calibration time, and at the same time provides quantitative, high-resolution images with a fast processing time.

Conveniently, DL models could also be an asset in handling the very large amounts of data that are inextricably linked with a theranostic platform. For example, DL models allow automatic feature extraction, i.e. selecting those representations of the data that contain the most information,  moreover they are able to reduce the dimensionality of the data, resulting in a faster operation of the platform and again identifying those data components of interest to the decision support algorithms \cite{Goodfellow-et-al-2016}. DL has also found its use in the denoising of datasets and in regularizing\cite{8253590}.  

A disadvantage of DL models is that they require sufficient ground truth data sets (MNP heating vs applied magnetic field, MNP location vs magnetic measurement signals, etc.), while generally this data in our field is still scarce. Recent advances towards overcoming this hurdle have been set for MPI with the introduction of the magnetic particle imaging data format (MDF)\cite{knopp2016mdf}, an open document standard for the storage of MPI data, and the OpenMPIData initiative\cite{knopp2020openmpidata}, which contains freely accessible MPI data of phantoms with specific challenges. Other MNP-based applications have not yet devoted attention to open data and standardisation, although improvements are underway among others in the context of European projects and ISO standards for standardising procedures in magnetic hyperthermia experiments and in other magnetic measurements \cite{schier2019european,wells2017standardisation}. For example, the European COST action RADIOMAG analysed the differences in calculated SLP/ILP-values among 21 laboratories across Europe for identical particle samples, which unraveled a lack of harmonization in heating characterization measurements, and highlighted the need for standardized, quantitative characterization techniques \cite{Wells2021Challenges}. Another issue with DL models are the concerns raised about interpretability and accountability, as they are trained on specific cases and present relevant shortcomings in terms of explainability, replicability, and, ultimately, trustworthiness. Especially in the medical environment of theranostic setups, aforementioned points are crucial. 

To overcome the aforementioned disadvantages of DL models, currently hindering the adoption of DL in our field, we propose \textit{hybrid} or so-called \textit{grey-box models} as a solution. These combine the extrapolation capabilities of physics-inspired models (robust over a large domain, albeit with a lower accuracy) with the regression capabilities of data-driven models (accurate on similar data as the training data). This way, incomplete models and incomplete data suffice to make accurate predictions or gain scientific insights. Different approaches of combining physics and DL can be found in literature. For example, \textit{physics-informed neural networks} constrain the structure of the DL model based on physical laws (e.g. conservation principles)\cite{RAISSI2019686}, while another class directly employs physics-based layers in a DL model\cite{jia2019physics} or, alternatively, the DL model is used in compensating prediction discrepancies of the used physics-based model\cite{liu2020physics}. Grey-box approaches were shown to typically require significantly less training data compared to black-box models and could reveal unknown dynamics and system parameters \cite{de2021neural,de2021physics}, allowing to improve currently existing models and e.g. find unknown nanoparticle dynamics. 

An approach, similar to the \textit{physics-informed neural networks} described above, but only working on a single measurement, i.e. not yet hybrid, is the so-called Deep Image Prior (DIP) \cite{lempitsky2018deep}. The DIP is specifically useful for solving ill-posed inverse problems \cite{dittmer2020regularization}. In DIP, the parameters of the neural network are determined based on this single measurement, with the aim that the results found using the network correspond to the minimization of a \textit{prior}. Hence, the regularization is encoded in the determined architecture of the network. The DIP approach was adapted by \citet{dittmer2020deep} to be used in MPI reconstructions, which showed superior or similar reconstruction quality compared to traditional MPI reconstruction methods \cite{dittmer2020deep}. Up until now, full hybrid techniques have not been introduced in our field and would be a game-changer,  significantly improving current imaging and application models. Moreover, hybrid systems provide ways to gather insights in the choices made by the black-box part of the model, increasing the interpretability of the the data-driven system and of the dynamics occurring in the system \cite{de2022prediction}.

\begin{figure}
\includegraphics[width=\columnwidth]{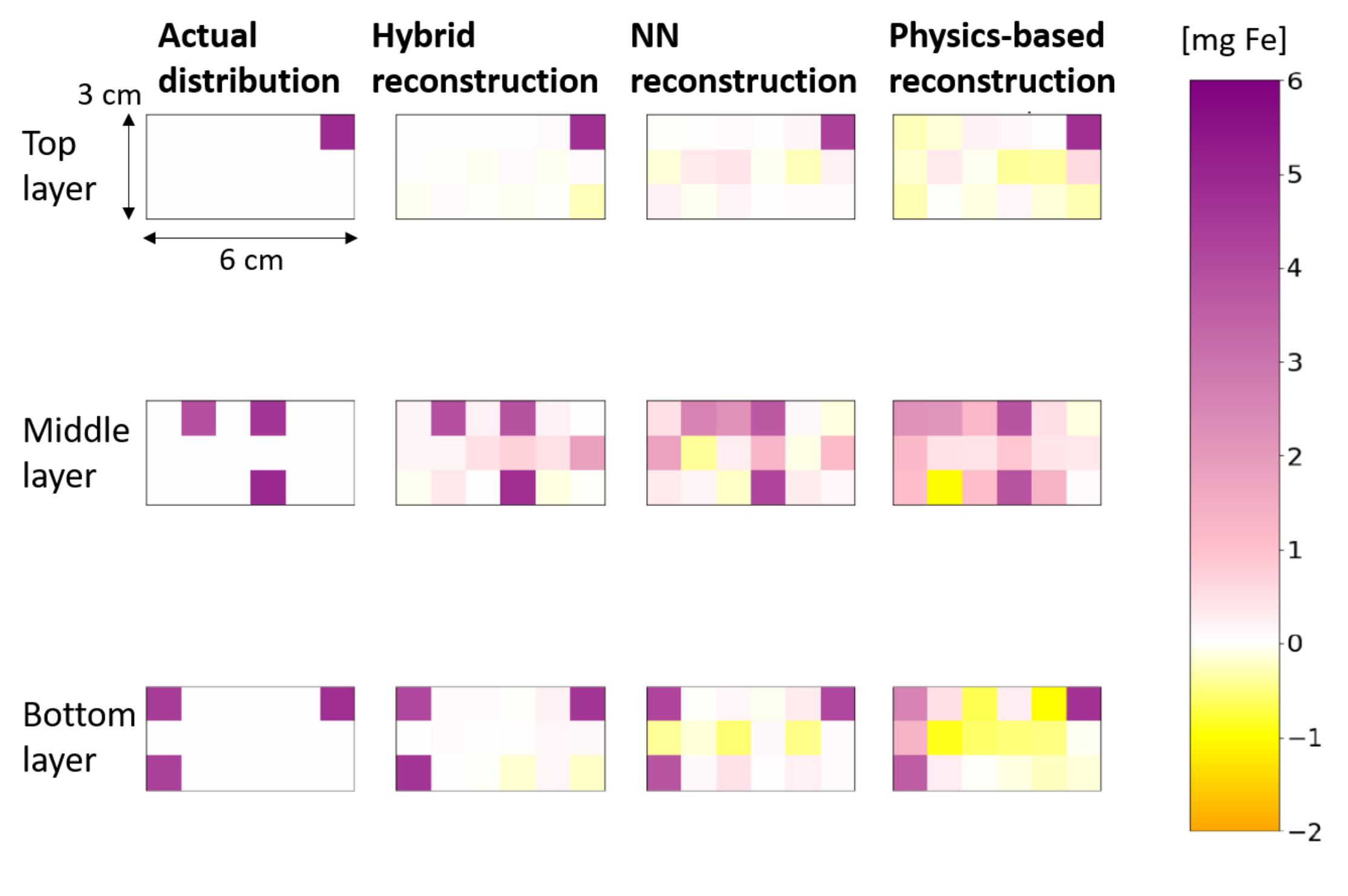}
\caption{\label{fig:hybridrecon} The imager described in Ref. \cite{liebl2014quantitative} allows sample sizes up to 3 cm (top, middle, bottom) $\times$ 6 cm $\times$ 3 cm. The physics-based model associated to the imager was employed to generate measurements of various MNP distributions in training and validation with added white Gaussian noise to realize signals with a SNR of 5 dB. A specific MNP distribution is shown (first column) with its associated reconstructions for a hybrid (second column), NN (third column) and physics-based (fourth column) approach.} 
\end{figure}

As a simple example to show the power of the hybrid approach, we combine a well-known physics-based model of an MRX imager \cite{liebl2014quantitative} with a fully connected NN \cite{Goodfellow-et-al-2016} and compare its performance in noisy simulations to using solely the NN without any physics knowledge and to using solely the physics-based MRXi model. The implementation was done in Keras \cite{chollet2021deep}. Measurements are generated for various MNP distributions using the physics-based model with added white Gaussian noise to achieve a SNR of 5 dB for both training (90000 samples) as well as validation (9000 samples). For the NN a single hidden layer is used and it has the generated measurements as input (2592 elements) with a spatial MNP distribution as output (54 elements). See table \ref{tab:table1} for the hyperparameters of the NN. Note that it is expected that more advanced DL configurations would have an increased impact. Especially the convolutional neural network as discussed before in case of imaging, more easily captures spatial relations and requires less parameters for a similar performance. This example rather serves to show that the combination of physics-based and data-driven models can enhance the performance compared to its stand-alone counterparts. In the hybrid model, the NN receives the input of the physics-based model, which partly relieves the NN from learning the underlying physics. Therefore, the input units of the NN in the hybrid approach are lowered to 2 $\times$ 54 as it receives the proposed MNP distribution from the physics-based model and the MNP distribution associated to the difference between the measurement and the measurement generated from the physics-based model. Figure \ref{fig:hybridrecon} shows a significant improvement in reconstruction quality when a hybrid approach is pursued, achieving a mean squared error (MSE) of 0.0012 between the actual MNP distribution and the hybrid reconstruction, whereas solely data-driven or physics-based reconstructions result in a MSE of 0.0022 and 0.015 respectively. Across the validation set a superior performance was seen for the hybrid approach with an average reduction in MSE of 29 \% compared to NN reconstructions and 49 \% to physics-based reconstructions. To explore the full extent of improvements that can be made using such aproaches certainly requires additional research using actual measurements and more complex DL techniques.  

\begin{table}[htbp]
  \centering
  \caption{Defined hyperparameters of the fully connected NN. SGD stands for stochastic gradient descent and ReLU for Rectified Linear Unit.}
    \begin{tabular}{|c|c|}
    \hline
    Input units (number of measurements) & 2592 \\
    \hline
    Hidden units & 874 \\
    \hline
    Dropout rate in hidden layer & 0.5 \\
    \hline
    Activation function & ReLU \\
    \hline
    Output units (number of voxels in sample) & 54 \\
    \hline
    Training entries & 90000 \\
    \hline
    Validation entries  & 9000 \\
    \hline
    Training epochs  & 200 \\
    \hline
    Learning rate  & 0.05 \\
    \hline
    Learning rate decay & Reduce by 20\%/epoch \\
    \hline
    Batch size & 100 \\
    \hline
    Optimizer & SGD with momentum \\
    \hline
    Noise level measurements & 5 dB \\
    \hline
    \end{tabular}%
  \label{tab:table1}%
\end{table}%

As data-driven and hybrid models are still rather unexplored territory in our field, this research track contains many uncertainties and risks, but, given the outstanding performance of data-driven alternatives compared to the traditional techniques in many fields like system identification and imaging, and the success of hybrid models on limited data sets, it also has a huge potential in improving the performance of a theranostic platform and in determining which data should be acted upon by the system.

\section{Conclusions}
In this perspective we highlighted open challenges and opportunities that could significantly advance MNP-based theranostics. Contrasting other literature, which rather focus on required MNP developments, here the underlying magnetic imaging technique with its associated applications are addressed as they form an important cornerstone of theranostic applications. First an overview was given of recent advances in MNP imaging techniques and, building upon this, how these techniques can be used in combination with MDT and magnetic hyperthermia as first steps towards theranostic applications. 

In order to continue progress in this field, we show the benefits of exploiting the rich magnetic dynamics of the MNPs by introducing novel magnetic field sequences, as this allows to increase the information content on the MNPs in the signals, while simultaneously reducing measurement time so to allow the practical realisation of theranostic platforms. Moreover, innovative combinations could also directly boost performance of e.g. magnetic hyperthermia and MDT applications by initialising or breaking up MNP clusters and modifying particle interactions. 

A second opportunity which might stimulate progress is the introduction of both imaging and application models in the theranostic platform that provide feedback to one another. We demonstrate how feedback can significantly increase the performance of applications. Care needs to be taken to select accurate and useful feedback to not jeopardize platform stability. In order to do so, while maintaining real-time information, interacting models of different complexity levels need to be developed. 

The first two aforementioned opportunities give rise to a significant increase in measurement and computational data, which opens up the pathway to the third opportunity, which is the use of DL models to handle the vast differences in data, and to perform automatic feature extraction and data dimensionality reduction. These properties aid in selecting those data streams which are most beneficial for the platform, while ensuring fast throughput of the data. Nevertheless, labeled data of the theranostic platform is still scarce, complicating the training of these techniques. One way to overcome this data requirement is the introduction of hybrid DL techniques in our field that combine physics-based as well as data-driven techniques with the additional advantages that unknown MNP dynamics can be retrieved and the data-driven part of the theranostic platform can be made interpretable. The latter is of paramount importance in a medical environment. We applied the hybrid approach on imaging data and showed a significant improvement in reconstruction quality compared to a completely data-driven reconstruction or a physics-based reconstruction.

As shown in this perspective, the endless tailoring possibilities of the MNPs together with the application of unlimited magnetic field configurations for exploiting their complex dynamics, supported by smart and hybrid models, prove a promising as well as challenging playground for further optimizing MNP-based theranostics and hence opening up the pathway of these systems towards clinical practice.

\section{Acknowledgements}
A. C. (12Z4722N) and J. L. (12W7622N)  are supported by the Research Foundation – Flanders (FWO) through senior postdoctoral research fellowships.

\appendix*

\section{List of abbreviations}
AC: alternating current\\
AMF: alternating magnetic field\\
CPU: central processing unit\\
CT: computed tomography\\
DC: direct current\\
DIP: deep image prior\\
DL: deep learning\\
FFR: field-free region\\
GPU: graphics processing unit\\
ILP: intrinsic loss power\\
MDF: magnetic particle imaging data format\\
MDT: magnetic drug targeting\\
MNP: magnetic nanoparticle\\
MPI: magnetic particle imaging\\
MRI: magnetic resonance imaging\\
MR: magnetic resonance\\
MRXi: magnetorelaxometry imaging\\
MRX: magnetorelaxometry\\
MSE: mean squared error\\
MSI: magnetic susceptibility imaging\\
NN: neural network\\
PET: positron emission tomography\\
RPM: Response and parameter mapping\\
SAR: specific absorption rate\\
SLP: specific loss power\\
SPECT: Single-photon emission computed tomography\\
US: ultrasound\\

\bibliography{aipsamp}
\end{document}